\def\cm{cm$^{\text{-1}}$}
\def\arc{$\alpha$-RuCl$_3$}

\def\supcondambient{S2}
\def\supddtrans{S3}

\def\supmagnet{S5}

\def\supcalc1{S9}
\def\supcalc2{S10}
\documentclass[aps,prl,twocolumn,superscriptaddress,showpacs,amsmath,amssymb]{revtex4}
\usepackage{graphicx}%
\usepackage{color}%
\usepackage{xspace}
\usepackage{ulem}
\usepackage{xcolor}

\begin{document}
\title{Detuning the Honeycomb of $\alpha$-RuCl$_{3}$:\\
Pressure-Dependent Optical Studies Reveal Broken Symmetry}
    \author{Tobias Biesner}
    \affiliation{1.~Physikalisches Institut, Universit\"at Stuttgart, Pfaffenwaldring 57, 70550 Stuttgart, Germany}
    \author{Sananda Biswas}
    \affiliation{Institut f\"{u}r Theoretische Physik, Goethe-Universit\"{a}t Frankfurt, 60438 Frankfurt am Main, Germany}
    \author{Weiwu Li}
    \affiliation{1.~Physikalisches Institut, Universit\"at Stuttgart, Pfaffenwaldring 57, 70550 Stuttgart, Germany}
    \author{Yohei Saito}
    \affiliation{1.~Physikalisches Institut, Universit\"at Stuttgart, Pfaffenwaldring 57, 70550 Stuttgart, Germany}
    \author{Andrej Pustogow}
    \affiliation{1.~Physikalisches Institut, Universit\"at Stuttgart, Pfaffenwaldring 57, 70550 Stuttgart, Germany}
    \author{Michaela Altmeyer}
    \affiliation{Institut f\"{u}r Theoretische Physik, Goethe-Universit\"{a}t Frankfurt, 60438 Frankfurt am Main, Germany}
    \author{Anja U. B. Wolter}
    \affiliation{Leibniz Institut f\"ur Festk\"orper- und Werkstoffforschung (IFW) Dresden, 01171 Dresden, Germany}
    \author{Bernd B\"uchner}
    \affiliation{Leibniz Institut f\"ur Festk\"orper- und Werkstoffforschung (IFW) Dresden, 01171 Dresden, Germany}
    \affiliation{Institut f\"ur Festk\"orperphysik, Technische Universit\"at Dresden, 01062 Dresden, Germany}
    \author{Maria Roslova}
    \affiliation{Fachrichtung Chemie und Lebensmittelchemie, Technische Universit\"at Dresden, 01062 Dresden, Germany}
    \author{Thomas Doert}
    \affiliation{Fachrichtung Chemie und Lebensmittelchemie, Technische Universit\"at Dresden, 01062 Dresden, Germany}
    \author{Stephen M. Winter}
    \affiliation{Institut f\"{u}r Theoretische Physik, Goethe-Universit\"{a}t Frankfurt, 60438 Frankfurt am Main, Germany}
    \author{Roser Valent\'{\i}}
    \affiliation{Institut f\"{u}r Theoretische Physik, Goethe-Universit\"{a}t Frankfurt, 60438 Frankfurt am Main, Germany}
    \author{Martin Dressel}
    \affiliation{1.~Physikalisches Institut, Universit\"at Stuttgart, Pfaffenwaldring 57, 70550 Stuttgart, Germany}

    \date{\today}

\begin{abstract}
The honeycomb Mott insulator $\alpha$-RuCl$_3$ loses its low-temperature magnetic order by pressure. We report clear evidence for a dimerized structure at $P>1$~GPa and observe the breakdown of the relativistic $j_{\rm eff}$ picture in this regime strongly affecting the electronic properties. A pressure-induced Kitaev quantum spin liquid cannot occur in this broken symmetry state. We shed light on the new phase by broad-band infrared spectroscopy of the low-temperature properties of $\alpha$-RuCl$_{3}$ and  {\it ab initio} density functional theory calculations, both under hydrostatic pressure.
\end{abstract}

\pacs{
75.10.Jm,    
75.10.Kt,    
78.20.-e,    
61.50.Ks,    
78.30.-j    
}
\maketitle

Interest in quantum spin liquids has grown steadily during the last decade, as more materials could be realized successfully. For instance, the organic charge transfer salts reach a high degree of frustration by forming an almost ideal triangular lattice \cite{Zhou17} or the herbertsmithites crystallize in a perfect kagome lattice \cite{Norman16}. In all of these examples no signs of magnetic order could be detected down to temperatures several orders of magnitude below the dominant magnetic couplings -- thus providing strong evidence for a quantum spin liquid state.

Over the last years a new class of frustrated magnets, the so-called honeycomb Kitaev systems~\cite{Jackeli2009,Trebst17,winter2017models,hermanns2017physics}, has attracted the attention of the condensed matter community. Prominent examples are \arc\ \cite{Plumb14,PhysRevB.92.235119} together with Li$_2$IrO$_3$ and Na$_2$IrO$_3$~\cite{winter2017models}. Although \arc\ shows zigzag antiferromagnetic order below $T_{\rm N}\approx 7$~K, neutron and Raman scattering experiments identified a continuum of excitations~\cite{Sanilands15,Banerjee16,Banerjee17} whose origin is intensively discussed in terms of magnon breakdown~\cite{Winter17a} and possible fractionalized Majorana excitations~\cite{Banerjee17}. Furthermore, \arc\ undergoes a transition to a quantum disordered state in an external magnetic field~\cite{Wolter17,Baek17} and suggestions of appearance of possible spin-liquid behavior are presently being discussed~\cite{PhysRevLett.117.277202,Winter17b}. The application of pressure to further tune the magnetic couplings is an attractive approach only recently being explored \cite{Wang17b,Cui17}.

Specific heat measurements  \cite{Wang17b} reveal that the N{\'e}el temperature $T_{\rm N}$ of \arc\ is initially enhanced by pressure, but magnetic order is sharply suppressed at a pressure of $P \approx 0.7$~GPa. NMR and magnetization studies consistently indicated a magnetically disordered high-pressure phase with strongly reduced susceptibility \cite{Cui17}; hence, Cui {\it et al.} posited the existence of a structural instability. This reminds of a recent investigation on $\alpha$-Li$_2$IrO$_3$ \cite{Hermann18}, which indicated dimerization above a critical pressure. Such instabilities could be rather general to the family of honeycomb Kitaev materials~\cite{Hermann18,streltsov2017orbital,Veiga17}. In order to shed light on these issues, we have conducted comprehensive spectroscopic investigations of the optical properties of \arc\ under pressure, which are combined with {\it ab initio} density functional theory calculations of the phonon spectrum and electronic properties. Our observations show that \arc\ undergoes a structural transition at moderate pressures where Ru-Ru dimers are formed.

High-quality single crystals of \arc\ were grown by chemical vapour transport as described in Ref.~\cite{Hentrich17}. The crystals were characterized by magnetic susceptibility measurements with the field along the $ab$--plane where magnetic order at $T_{\rm N} \approx 11$~K and 8~K is observed; the high temperature Curie-Weiss fit yields $\theta_{\rm CW}=38$~K and $\mu_{\rm eff}=2.3~\mu_B$. Optical reflectivity experiments employed several Fourier-transform spectrometers covering the range from 100 up to 20\,000~\cm. For measurements from the near-infrared up to the ultraviolet spectral range spectroscopic ellipsometry was utilized. Ambient pressure experiments were performed in helium-bath and flow cryostats, with the help of an infrared microscope if required. While the magnetic transitions at $T_{\rm N}$ do show up as minor changes in the optical spectra, applying an external magnetic field up to 7.4~T (both in-plane and out-of-plane) does not affect the infrared transmission noticeably. In addition, we conducted reflectivity measurements in a piston pressure cell operating up to 2~GPa and down to temperatures as low as 10~K  by using Daphne oil as the pressure transmitting medium \cite{Beyer15}. For those measurements we prepared a powder and pressed pellets. The optical conductivity was obtained from a Kramers-Kronig analysis of the combined data using common high- and low-frequency extrapolations.

\begin{figure}[t]
\centering
\includegraphics[width=1\columnwidth]{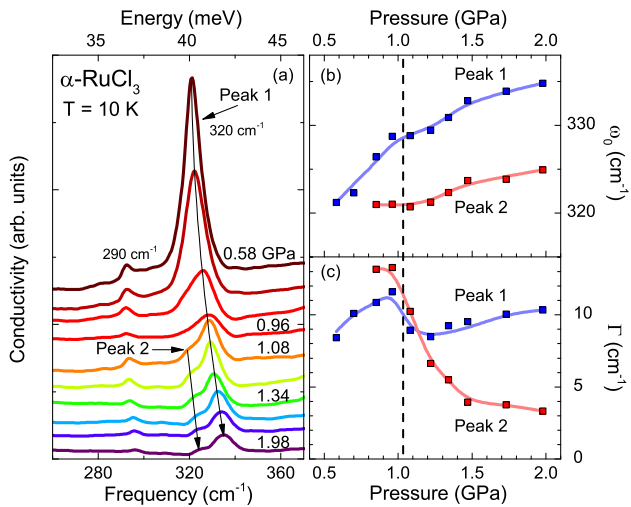}
\caption{(a)~Low-temperature conductivity spectra in the range of the 290 and 320~\cm\ phonon modes recorded for different hydrostatic pressure. The values refer to the pressure set at room temperature. (b)~Position $\omega_0$ and (c) linewidth $\Gamma$ of peak 1 and peak 2 obtained from fits of the low-temperature conductivity for different pressure applied. The colored lines are guides to the eye.
\label{fig:phonon}}
\end{figure}

\begin{figure}[!t]
\centering
\includegraphics[width=0.9\columnwidth]{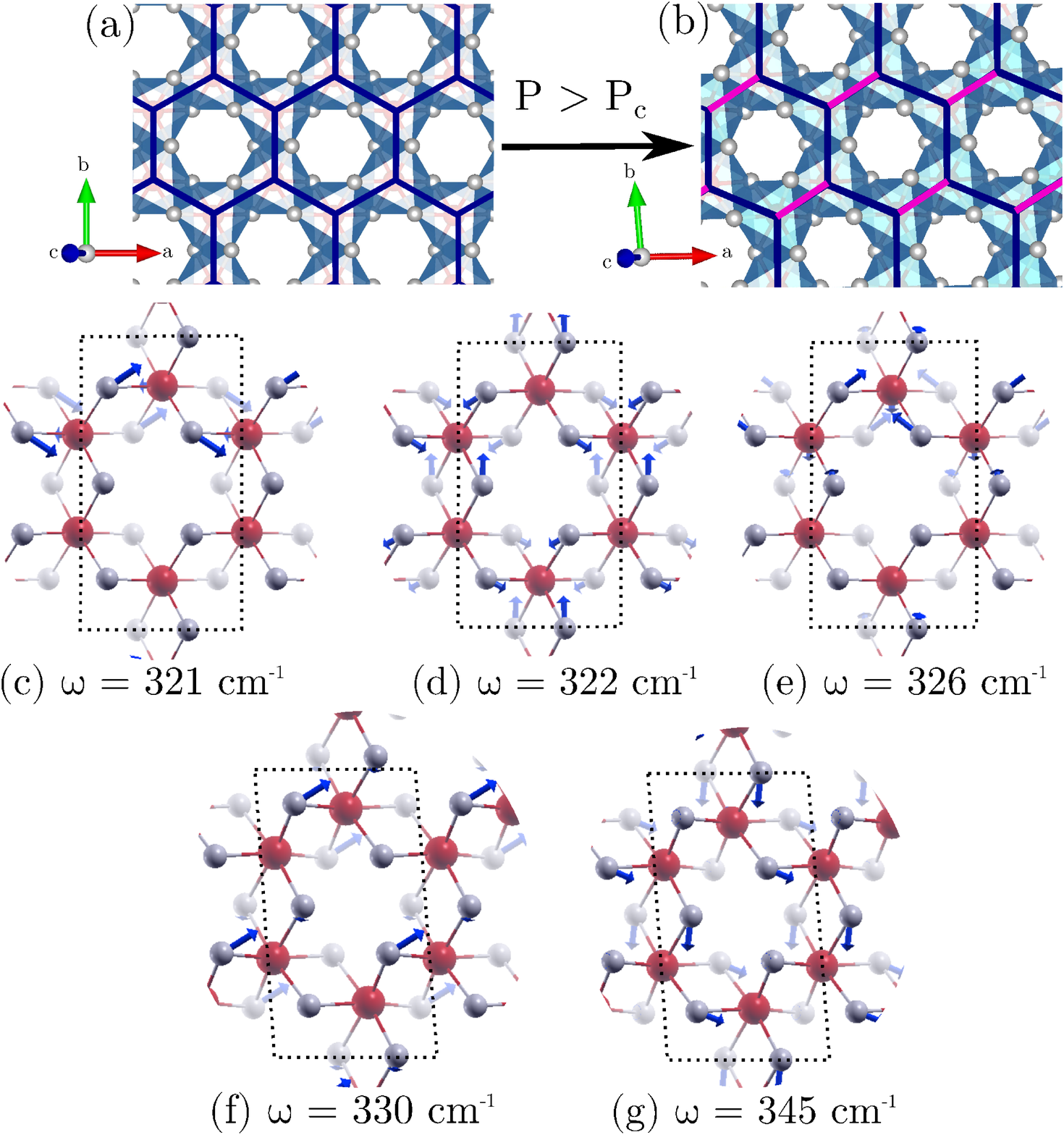}
\caption {(a)--(b) Top views of the $ab$--plane in undimerized C2/$m$ and dimerized P$\bar{1}$ structures, respectively; the dimerization is indicated by magenta lines. The corresponding calculated in-plane phonon modes (blue arrows) and frequencies are presented in (c)--(e) for $P < P_{c}$ and in (f)--(g) for $P > P_{c}$. Ru and Cl atoms are denoted by red and grey spheres, respectively.
\label{fig:2}}
\end{figure}

In order to identify a possible structural transition under pressure, we begin with the phononic contributions to the optical spectra of
\arc, observed below 400~\cm.
In Fig.~\ref{fig:phonon}(a) the optical conductivity at $T=10$~K is displayed for the whole pressure range. Two phonon modes are clearly visible at 290~\cm\ and 320~\cm\ (peak 1) at low pressures. In previous ambient-pressure infrared \cite{Hasegawa17} and Raman experiments \cite{Glamazda17}, these two features were identified as out-of-plane $A_{2u}$ and in-plane $E_u$ vibrations, respectively, with reference to an idealized P$\bar{3}1m$ symmetry of the individual RuCl$_3$ layers.
For the presently studied powder samples, the presence of stacking faults likely reduces the symmetry even at low pressure, effectively mixing the ambient pressure ABC (rhombohedral R$\bar{3}$) and AB (monoclinic C2/$m$) stacking motifs. In order to assign the vibrations, we therefore performed phonon calculations using the linear response method~\cite{Giannozzi09,Baroni01} in the lower C2/$m$ symmetry (see Fig.~\ref{fig:2}). The displacements are sketched in Fig.~\ref{fig:2}(c--g). At ambient pressure, we find a weak infrared-active out-of-plane mode with frequency 287~\cm, which can be identified with the observed peak at 290~\cm\ (see Supplemental Material \cite{SM}). At higher frequencies, the calculations suggest a trio of nearly degenerate modes with mostly in-plane polarization vectors; the frequencies of these modes are computed to be $\omega_0 \approx 321$, 322, and 326~\cm. The near-degeneracy of the modes ($\Delta \omega/\omega_0 \approx 1.5$\%) stems from the quasi-threefold symmetry of the individual RuCl$_3$ layers, which is preserved regardless of the stacking pattern.  The pronounced deflection of Ru and Cl sites for the modes near 320~\cm\ provides a large dipole moment, resulting in a larger intensity compared to the 290~\cm\ vibration. Complementary optical transmission measurements on single crystals at ambient pressure support these assignments; they are presented in Fig.~\supmagnet~\cite{SM}.

Under applied pressure, both modes shift to high energy as expected for reduced lattice spacing. For $P > 0.7$~GPa, the 320~\cm\ peak (1) develops a pronounced shoulder (peak 2 in Fig.~\ref{fig:phonon}(a)), which eventually splits off with increasing $P$, producing two features with an intensity ratio of approximately 1:2. In Fig.~\ref{fig:phonon}(b) and (c) the frequency shift and linewidth $\Gamma$ are quantified as a function of pressure, on the basis of Lorentzian fits. Between 0.96 and 1.08~GPa ($P_c$) the linewidth of the main mode around $\omega_0 = 320$~\cm\ suddenly decreases to $\Gamma \approx 10$ \cm\ while the phonon mode splits. This large splitting can be taken as direct evidence that the quasi-threefold lattice symmetry of the layers is broken at high pressure, suggesting the possibility of dimerization.

In order to verify whether \arc\ is susceptible to dimerization under pressure, we performed {\it ab initio} calculations for structural optimization under pressure (see Supplemental Materials \cite{SM} for all computational details).
Similar to previous calculations on $\alpha$-Li$_2$IrO$_3$~\cite{Hermann18}, we find that parallel dimerization becomes energetically favored at high pressure over the homogeneous ambient pressure structure. The high-pressure phase is triclinic P$\bar{1}$, and is shown in Fig.~\ref{fig:2}(b); the computation of phonons for the dimerized structure shows a splitting of the almost degenerate modes in the homogeneous structure [Fig.~\ref{fig:2}(c--e)] into two dominant in-plane modes [Fig.~\ref{fig:2}(f,g)] with frequencies reaching 330 and 345~\cm\ in the limit $P = 10$ GPa. Consistent with the experiment, we also find that the calculated infrared intensity of the latter mode is higher than the former. The large computed splitting of $\Delta \omega\approx 15$~\cm\ compares well with the results from the conductivity measurements in Fig.~\ref{fig:phonon}(b), providing further support for a pressure-driven structural phase transition of \arc\ to a dimerized phase.

\begin{figure}
\centering
\includegraphics[width=1\columnwidth]{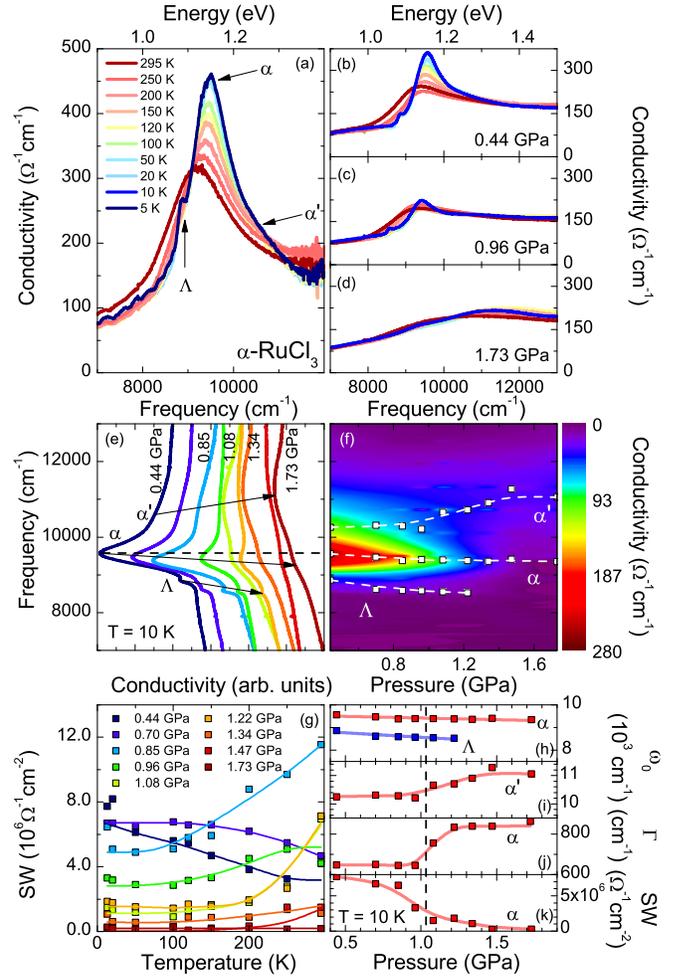}
\caption{(a)~Temperature dependence of the conductivity spectrum of \arc\ around the Mott gap. Three distinct features are indicated by arrows: the small excitonic peak $\Lambda$, the main absorption maximum $\alpha$, and the shoulder $\alpha^{\prime}$. \text{(b--d)}~Temperature evolution of the optical conductivity for selected
hydrostatic pressure as indicated. (e)~Pressure dependence of the low temperature optical conductivity. For clarity reasons the curves are shifted with respect to each other. The pressure effects can be followed in the false-color contour plot (f). In panel (g)~the spectral weight (SW) is displayed as a function of temperature. The parameters are obtained from fits of the $\alpha$-feature measured at different pressure values as indicated. Pressure dependence of (h,i)~the positions of the
$\alpha$-, $\Lambda$-, and $\alpha^{\prime}$-peaks, respectively; (j)~the linewidth $\Gamma$ of the $\alpha$-peak, (k)~the spectral weight, obtained from
fits of the low-temperature conductivity for different pressure applied. The colored lines are guides to the eye.
\label{fig:condTp}}
\end{figure}

This observed reduction of lattice symmetry is expected to severely affect the electronic structure.
In order to determine the electronic properties under pressure, we performed temperature-dependent infrared reflectivity experiments on \arc\ pellets while applying hydrostatic pressures step-by-step up to 1.7~GPa. A general overview of the electrodynamic properties is given in Fig.~\ref{fig:condTp}(a). The sharp absorption edge at 1.1~eV ($\alpha$-peak) is assigned to intersite $d^5$-$d^5 \to d^6$-$d^4$ excitations \cite{Fletcher67,Binotto71,Guizzetti79} yielding a final $d^4$ triplet $^3T_1$ state. Transitions to higher-energy multiplets appear above 1.5 eV. In addition to the main $\alpha$-peak, on the high-frequency wing a weak shoulder appears around 10\,500~\cm, which we call $\alpha^{\prime}$. On the lower-frequency side, a narrow peak
labeled $\Lambda$ develops at 8800~\cm\ below $T=120$~K, which is likely a bound excitonic state \cite{Sandilands16b}. Intrasite $t_{2g}^5$-$e_g^0 \to t_{2g}^4$-$e_g^1$ excitations (see Fig.~\supddtrans~\cite{SM}) are also identified at 0.28~eV \cite{Guizzetti79,Sandilands16a}, while higher energy bands around 5.2~eV originate from transitions between Cl $3p$ and Ru $4d$ states [Fig.~\supcondambient~\cite{SM}].

Fig.~\ref{fig:condTp}(b--d) displays the temperature evolution of the conductivity spectra for selected pressure values. The effect of $P$ on the low-temperature conductivity can be followed in Fig.~\ref{fig:condTp}(e,f). From Lorentz fits of the conductivity we obtain the pressure-dependent peak position, spectral weight and width for each contribution, as plotted in Fig.~\ref{fig:condTp}(g--k). The $\alpha$- and $\Lambda$-features generally broaden and slightly redshift with increasing $P$. In accord with transport data \cite{Wang17b}, our optical measurements therefore do not reveal a closure of the Mott gap under pressure. Near $P_c$, the spectral weight of the $\alpha$-feature instead diminishes dramatically by almost one order of magnitude, cf.\ Fig.~\ref{fig:condTp}(k). The drop is most pronounced between $P=0.85$ and 1.08~GPa. When pressurized further, the $\alpha$-peak cannot be well identified anymore.
The excitonic $\Lambda$-feature behaves likewise before it vanishes above 1.22~GPa. The similar pressure dependence observed for both peaks $\alpha$ and $\Lambda$ [Fig.~\ref{fig:condTp}(h)] not only supports the assignment of $\Lambda$ as an excitonic resonance \cite{Sandilands16b} but sheds light on their common physical origin. From the $\alpha^{\prime}$-shoulder a broad maximum emerges around 11\,000~\cm\ above $P_c$, which becomes more pronounced with cooling and eventually dominates the entire infrared spectrum.  This mode rapidly shifts up by about 800~\cm\ in the range from 0.44 to 1.73~GPa, as shown in Fig.~\ref{fig:condTp}(i). This distinct pressure dependence suggests a different physical origin compared to the $\alpha$ and $\Lambda$ modes.

Focusing on the $\alpha$ peak, Fig.~\ref{fig:condTp}(g) displays the $T$ dependence of the spectral weight (SW) obtained from Lorentz fits. Up to $P=0.7$~GPa, the $\alpha$-peak notably intensifies upon cooling. Following \cite{Koitzsch17b,Khaliullin04}, the intensity of such excitations to the $^3T_1$ triplet state is expected to scale like $I\propto  1+ 4 \langle S_i^\gamma S_j^\gamma\rangle$ ($\gamma =x,y,z$), providing a local probe of the magnetic correlations. An analogous enhancement of the $\alpha$-peak intensity on cooling in recent EELS measurements was thus ascribed to the development of short-ranged ferromagnetic correlations (i.e. $ \langle S_i^\gamma S_j^\gamma\rangle > 0$) below $T= 100$~K \cite{Koitzsch17b}. At higher pressures, we find that the spectral weight behaves fundamentally differently. For $P \geq 0.85$~GPa it {\it decreases} upon cooling down to $T=50$~K and saturates at lower temperatures. For $P > 1.22$~GPa the spectral weight is almost completely suppressed, and nearly independent on $T$. The suppression of the $\alpha$-peak therefore strongly suggests a collapse of the ferromagnetic Kitaev interactions above $\sim$ 0.7 GPa, which may be related to (i) a breakdown of the $j_{\rm eff}$ picture, and/or (ii) the development of strong intradimer antiferromagnetic correlations. The latter effect is consistent with the greatly suppressed magnetic susceptibility observed above 0.7~GPa \cite{Cui17}.

\begin{figure}
\centering
\includegraphics[width=\columnwidth]{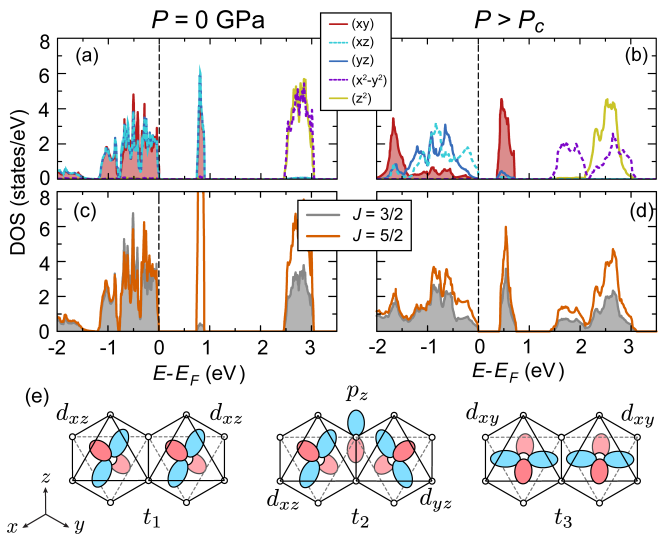}
\caption{\label{fig:dos}(a)--(d) Calculated density of states (DOS) using the GGA+SOC+$U$ method with $U=1.5$ eV at ambient pressure (left panels) and $P>P_c$ (right panels). The 
orbital-decomposed (top) and $J$--decomposed (bottom) DOS are compared in the two pressure regimes. Note that the $J=5/2$ and $J=3/2$  have most contributions coming from $j_{\rm eff}=1/2$ and $j_{\rm eff}=3/2$,  respectively.
(e) Schematics of Ru-Ru hopping pathways.}
\end{figure}

In order to gain insight into the high-pressure electronic structure, in Fig.~\ref{fig:dos}(a--d) we show the calculated orbital-dependent density of states (DOS) within the GGA+SOC+$U$ method and the FPLO basis with $U=1.5$ eV for the undimerized ($P=0$) and dimerized $P>P_c$ structures. For $P=0$ GPa, the narrow peak in the DOS at $\sim 1$~eV represents the single $t_{2g}$ hole at each Ru site, which resides in a relativistic $j_{\rm eff} = 1/2$ state composed of nearly equal contributions from each of the $d_{xy}, d_{xz}$, and $d_{yz}$ orbitals [Fig.~\ref{fig:dos}(a)]. The assignment is further verified by projecting onto the atomic $J$-states, for which the $j_{\rm eff} = 1/2$ orbital has purely $J = 5/2$ contributions, as shown in  Fig.~\ref{fig:dos}(c). In this case, the hopping of holes between Ru sites occurs largely through Cl $p_z$ orbitals, as described by the hopping integral $t_2$, shown in Fig.~\ref{fig:dos}(e). This leads to a dominant ferromagnetic Kitaev interaction $K S_i^\gamma S_j^\gamma$ at low $P$ that scales as $K \propto -J_H(t_2)^2/U^2$ \cite{Jackeli2009}, although other magnetic interactions of similar scale are also present due to finite $t_1$ and $t_3$ \cite{Rau14,Winter16}.

In contrast, for $P > P_c$, the dimerization is manifested by a strong splitting of the $d_{xy}$ orbitals into bonding and antibonding states as shown in Fig.~\ref{fig:dos}(b), which destroys the relativistic $j_{\rm eff}$ states [compare to Fig.~\ref{fig:dos}(d)]. These effects arise from a strong enhancement of the direct Ru-Ru $t_3$ hopping path along the dimerized bonds. Within the dimers, the magnetic couplings are thus dominated by large anti\-ferromagnetic Heisenberg $J \ \mathbf{S}_i \cdot \mathbf{S}_j$ interactions, with $J\propto +(t_3)^2/U$  \cite{Rau14,Winter16}.
Consistently, dimerization leads to complete suppression of the computed magnetic moments for the high pressure $P\bar{1}$ structure at the level of GGA+SOC+$U$.
This observation explains the suppression of zig-zag order at pressures around 1~GPa \cite{Cui17,Wang17b}, in favour of a gapped state. These findings are all in analogy with the honeycomb iridate $\alpha$-Li$_2$IrO$_3$ \cite{Hermann18}.

We conclude from our pressure-dependent optical investigation of the electronic and phononic properties of \arc\ that the structural symmetry is broken around 1~GPa. The dimerization of Ru-Ru bonds at high pressure is seen by the splitting of the 320~\cm\ phonon modes, which is consistently observed in the DFT calculations. This dimerization has dramatic effects on the electronic structure, as evidenced by the disappearance of the excitonic $\Lambda$ feature above 1~GPa and the development of the second $\alpha^\prime $ excitation on the high energy side of the optical gap. The temperature and pressure dependence of the optical response suggest a collapse of the Kitaev couplings in the dimerized phase, which is further motivated by theoretical analysis of the electronic structure. These observations rule out the possibility that \arc\ can be tuned towards a Kitaev spin liquid state under pressure. Instead it appears as a gapped dimerized phase.

\begin{acknowledgements}
We thank Ying Li and Kira Riedl for fruitful discussions; Micha Schilling for assistance in the magneto-optical measurements
and Gabriele Untereiner for continuous experimental support. The project was supported by the Deutsche Forschungsgemeinschaft (DFG) through grant SFB/TR~49, SFB 1143, DR228/39-1 and DR228/51-1.
\end{acknowledgements}

\bibliography{RuCl}

\begin{thebibliography}{37}
\expandafter\ifx\csname natexlab\endcsname\relax\def\natexlab#1{#1}\fi
\expandafter\ifx\csname bibnamefont\endcsname\relax
  \def\bibnamefont#1{#1}\fi
\expandafter\ifx\csname bibfnamefont\endcsname\relax
  \def\bibfnamefont#1{#1}\fi
\expandafter\ifx\csname citenamefont\endcsname\relax
  \def\citenamefont#1{#1}\fi
\expandafter\ifx\csname url\endcsname\relax
  \def\url#1{\texttt{#1}}\fi
\expandafter\ifx\csname urlprefix\endcsname\relax\def\urlprefix{URL }\fi
\providecommand{\bibinfo}[2]{#2}
\providecommand{\eprint}[2][]{\url{#2}}

\bibitem[{\citenamefont{Zhou et~al.}(2017)\citenamefont{Zhou, Kanoda, and
  Ng}}]{Zhou17}
\bibinfo{author}{\bibfnamefont{Y.}~\bibnamefont{Zhou}},
  \bibinfo{author}{\bibfnamefont{K.}~\bibnamefont{Kanoda}}, \bibnamefont{and}
  \bibinfo{author}{\bibfnamefont{T.-K.} \bibnamefont{Ng}},
  \bibinfo{journal}{Rev. Mod. Phys.} \textbf{\bibinfo{volume}{89}},
  \bibinfo{pages}{025003} (\bibinfo{year}{2017}).

\bibitem[{\citenamefont{Norman}(2016)}]{Norman16}
\bibinfo{author}{\bibfnamefont{M.~R.} \bibnamefont{Norman}},
  \bibinfo{journal}{Rev. Mod. Phys.} \textbf{\bibinfo{volume}{88}},
  \bibinfo{pages}{041002} (\bibinfo{year}{2016}).

\bibitem[{\citenamefont{Jackeli and Khaliullin}(2009)}]{Jackeli2009}
\bibinfo{author}{\bibfnamefont{G.}~\bibnamefont{Jackeli}} \bibnamefont{and}
  \bibinfo{author}{\bibfnamefont{G.}~\bibnamefont{Khaliullin}},
  \bibinfo{journal}{Phys. Rev. Lett.} \textbf{\bibinfo{volume}{102}},
  \bibinfo{pages}{017205} (\bibinfo{year}{2009}).

\bibitem[{\citenamefont{Trebst}(2017)}]{Trebst17}
\bibinfo{author}{\bibfnamefont{S.}~\bibnamefont{Trebst}},
  \bibinfo{journal}{arXiv:1701.07056}  (\bibinfo{year}{2017}).

\bibitem[{\citenamefont{Winter et~al.}(2017{\natexlab{a}})\citenamefont{Winter,
  Tsirlin, Daghofer, van~den Brink, Singh, Gegenwart, and
  Valent{\'{i}}}}]{winter2017models}
\bibinfo{author}{\bibfnamefont{S.~M.} \bibnamefont{Winter}},
  \bibinfo{author}{\bibfnamefont{A.~A.} \bibnamefont{Tsirlin}},
  \bibinfo{author}{\bibfnamefont{M.}~\bibnamefont{Daghofer}},
  \bibinfo{author}{\bibfnamefont{J.}~\bibnamefont{van~den Brink}},
  \bibinfo{author}{\bibfnamefont{Y.}~\bibnamefont{Singh}},
  \bibinfo{author}{\bibfnamefont{P.}~\bibnamefont{Gegenwart}},
  \bibnamefont{and}
  \bibinfo{author}{\bibfnamefont{R.}~\bibnamefont{Valent{\'{i}}}},
  \bibinfo{journal}{J. Phys. Condens. Matter} \textbf{\bibinfo{volume}{29}},
  \bibinfo{pages}{493002} (\bibinfo{year}{2017}{\natexlab{a}}).

\bibitem[{\citenamefont{Hermanns et~al.}(2018)\citenamefont{Hermanns, Kimchi,
  and Knolle}}]{hermanns2017physics}
\bibinfo{author}{\bibfnamefont{M.}~\bibnamefont{Hermanns}},
  \bibinfo{author}{\bibfnamefont{I.}~\bibnamefont{Kimchi}}, \bibnamefont{and}
  \bibinfo{author}{\bibfnamefont{J.}~\bibnamefont{Knolle}},
  \bibinfo{journal}{Annu. Rev. Condens. Matter Phys.}
  \textbf{\bibinfo{volume}{9}} (\bibinfo{year}{2018}).

\bibitem[{\citenamefont{Plumb et~al.}(2014)\citenamefont{Plumb, Clancy,
  Sandilands, Shankar, Hu, Burch, Kee, and Kim}}]{Plumb14}
\bibinfo{author}{\bibfnamefont{K.~W.} \bibnamefont{Plumb}},
  \bibinfo{author}{\bibfnamefont{J.~P.} \bibnamefont{Clancy}},
  \bibinfo{author}{\bibfnamefont{L.~J.} \bibnamefont{Sandilands}},
  \bibinfo{author}{\bibfnamefont{V.~V.} \bibnamefont{Shankar}},
  \bibinfo{author}{\bibfnamefont{Y.~F.} \bibnamefont{Hu}},
  \bibinfo{author}{\bibfnamefont{K.~S.} \bibnamefont{Burch}},
  \bibinfo{author}{\bibfnamefont{H.-Y.} \bibnamefont{Kee}}, \bibnamefont{and}
  \bibinfo{author}{\bibfnamefont{Y.-J.} \bibnamefont{Kim}},
  \bibinfo{journal}{Phys. Rev. B} \textbf{\bibinfo{volume}{90}},
  \bibinfo{pages}{041112} (\bibinfo{year}{2014}).

\bibitem[{\citenamefont{Johnson et~al.}(2015)\citenamefont{Johnson, Williams,
  Haghighirad, Singleton, Zapf, Manuel, Mazin, Li, Jeschke, Valent{\'{i}}
  et~al.}}]{PhysRevB.92.235119}
\bibinfo{author}{\bibfnamefont{R.~D.} \bibnamefont{Johnson}},
  \bibinfo{author}{\bibfnamefont{S.~C.} \bibnamefont{Williams}},
  \bibinfo{author}{\bibfnamefont{A.~A.} \bibnamefont{Haghighirad}},
  \bibinfo{author}{\bibfnamefont{J.}~\bibnamefont{Singleton}},
  \bibinfo{author}{\bibfnamefont{V.}~\bibnamefont{Zapf}},
  \bibinfo{author}{\bibfnamefont{P.}~\bibnamefont{Manuel}},
  \bibinfo{author}{\bibfnamefont{I.~I.} \bibnamefont{Mazin}},
  \bibinfo{author}{\bibfnamefont{Y.}~\bibnamefont{Li}},
  \bibinfo{author}{\bibfnamefont{H.~O.} \bibnamefont{Jeschke}},
  \bibinfo{author}{\bibfnamefont{R.}~\bibnamefont{Valent{\'{i}}}},
  \bibnamefont{et~al.}, \bibinfo{journal}{Phys. Rev. B}
  \textbf{\bibinfo{volume}{92}}, \bibinfo{pages}{235119}
  (\bibinfo{year}{2015}).

\bibitem[{\citenamefont{Sandilands et~al.}(2015)\citenamefont{Sandilands, Tian,
  Plumb, Kim, and Burch}}]{Sanilands15}
\bibinfo{author}{\bibfnamefont{L.~J.} \bibnamefont{Sandilands}},
  \bibinfo{author}{\bibfnamefont{Y.}~\bibnamefont{Tian}},
  \bibinfo{author}{\bibfnamefont{K.~W.} \bibnamefont{Plumb}},
  \bibinfo{author}{\bibfnamefont{Y.-J.} \bibnamefont{Kim}}, \bibnamefont{and}
  \bibinfo{author}{\bibfnamefont{K.~S.} \bibnamefont{Burch}},
  \bibinfo{journal}{Phys. Rev. Lett.} \textbf{\bibinfo{volume}{114}},
  \bibinfo{pages}{147201} (\bibinfo{year}{2015}).

\bibitem[{\citenamefont{Banerjee et~al.}(2016)\citenamefont{Banerjee, Bridges,
  Yan, Aczel, Li, Stone, Granroth, Lumsden, Yiu, Knolle et~al.}}]{Banerjee16}
\bibinfo{author}{\bibfnamefont{A.}~\bibnamefont{Banerjee}},
  \bibinfo{author}{\bibfnamefont{C.~A.} \bibnamefont{Bridges}},
  \bibinfo{author}{\bibfnamefont{J.-Q.} \bibnamefont{Yan}},
  \bibinfo{author}{\bibfnamefont{A.~A.} \bibnamefont{Aczel}},
  \bibinfo{author}{\bibfnamefont{L.}~\bibnamefont{Li}},
  \bibinfo{author}{\bibfnamefont{M.~B.} \bibnamefont{Stone}},
  \bibinfo{author}{\bibfnamefont{G.~E.} \bibnamefont{Granroth}},
  \bibinfo{author}{\bibfnamefont{M.~D.} \bibnamefont{Lumsden}},
  \bibinfo{author}{\bibfnamefont{Y.}~\bibnamefont{Yiu}},
  \bibinfo{author}{\bibfnamefont{J.}~\bibnamefont{Knolle}},
  \bibnamefont{et~al.}, \bibinfo{journal}{Nat. Mater.}
  \textbf{\bibinfo{volume}{15}}, \bibinfo{pages}{733} (\bibinfo{year}{2016}).

\bibitem[{\citenamefont{Banerjee et~al.}(2017)\citenamefont{Banerjee,
  Lampen-Kelley, Knolle, Balz, Aczel, Winn, Liu, Pajerowski, Yan, Bridges
  et~al.}}]{Banerjee17}
\bibinfo{author}{\bibfnamefont{A.}~\bibnamefont{Banerjee}},
  \bibinfo{author}{\bibfnamefont{P.}~\bibnamefont{Lampen-Kelley}},
  \bibinfo{author}{\bibfnamefont{J.}~\bibnamefont{Knolle}},
  \bibinfo{author}{\bibfnamefont{C.}~\bibnamefont{Balz}},
  \bibinfo{author}{\bibfnamefont{A.~A.} \bibnamefont{Aczel}},
  \bibinfo{author}{\bibfnamefont{B.}~\bibnamefont{Winn}},
  \bibinfo{author}{\bibfnamefont{Y.}~\bibnamefont{Liu}},
  \bibinfo{author}{\bibfnamefont{D.}~\bibnamefont{Pajerowski}},
  \bibinfo{author}{\bibfnamefont{J.~Q.} \bibnamefont{Yan}},
  \bibinfo{author}{\bibfnamefont{C.~A.} \bibnamefont{Bridges}},
  \bibnamefont{et~al.}, \bibinfo{journal}{arXiv:1706.07003}
  (\bibinfo{year}{2017}).

\bibitem[{\citenamefont{Winter et~al.}(2017{\natexlab{b}})\citenamefont{Winter,
  Riedl, Maksimov, Chernyshev, Honecker, and Valent{\'{i}}}}]{Winter17a}
\bibinfo{author}{\bibfnamefont{S.~M.} \bibnamefont{Winter}},
  \bibinfo{author}{\bibfnamefont{K.}~\bibnamefont{Riedl}},
  \bibinfo{author}{\bibfnamefont{P.~A.} \bibnamefont{Maksimov}},
  \bibinfo{author}{\bibfnamefont{A.~L.} \bibnamefont{Chernyshev}},
  \bibinfo{author}{\bibfnamefont{A.}~\bibnamefont{Honecker}}, \bibnamefont{and}
  \bibinfo{author}{\bibfnamefont{R.}~\bibnamefont{Valent{\'{i}}}},
  \bibinfo{journal}{Nat. Commun.} \textbf{\bibinfo{volume}{8}},
  \bibinfo{pages}{1152} (\bibinfo{year}{2017}{\natexlab{b}}).

\bibitem[{\citenamefont{Wolter et~al.}(2017)\citenamefont{Wolter, Corredor,
  Janssen, Nenkov, Sch{\"{o}}necker, Do, Choi, Albrecht, Hunger, Doert
  et~al.}}]{Wolter17}
\bibinfo{author}{\bibfnamefont{A.~U.~B.} \bibnamefont{Wolter}},
  \bibinfo{author}{\bibfnamefont{L.~T.} \bibnamefont{Corredor}},
  \bibinfo{author}{\bibfnamefont{L.}~\bibnamefont{Janssen}},
  \bibinfo{author}{\bibfnamefont{K.}~\bibnamefont{Nenkov}},
  \bibinfo{author}{\bibfnamefont{S.}~\bibnamefont{Sch{\"{o}}necker}},
  \bibinfo{author}{\bibfnamefont{S.-H.} \bibnamefont{Do}},
  \bibinfo{author}{\bibfnamefont{K.-Y.} \bibnamefont{Choi}},
  \bibinfo{author}{\bibfnamefont{R.}~\bibnamefont{Albrecht}},
  \bibinfo{author}{\bibfnamefont{J.}~\bibnamefont{Hunger}},
  \bibinfo{author}{\bibfnamefont{T.}~\bibnamefont{Doert}},
  \bibnamefont{et~al.}, \bibinfo{journal}{Phys. Rev. B}
  \textbf{\bibinfo{volume}{96}}, \bibinfo{pages}{041405}
  (\bibinfo{year}{2017}).

\bibitem[{\citenamefont{Baek et~al.}(2017)\citenamefont{Baek, Do, Choi, Kwon,
  Wolter, Nishimoto, van~den Brink, and B{\"{u}}chner}}]{Baek17}
\bibinfo{author}{\bibfnamefont{S.-H.} \bibnamefont{Baek}},
  \bibinfo{author}{\bibfnamefont{S.-H.} \bibnamefont{Do}},
  \bibinfo{author}{\bibfnamefont{K.-Y.} \bibnamefont{Choi}},
  \bibinfo{author}{\bibfnamefont{Y.~S.} \bibnamefont{Kwon}},
  \bibinfo{author}{\bibfnamefont{A.~U.~B.} \bibnamefont{Wolter}},
  \bibinfo{author}{\bibfnamefont{S.}~\bibnamefont{Nishimoto}},
  \bibinfo{author}{\bibfnamefont{J.}~\bibnamefont{van~den Brink}},
  \bibnamefont{and}
  \bibinfo{author}{\bibfnamefont{B.}~\bibnamefont{B{\"{u}}chner}},
  \bibinfo{journal}{Phys. Rev. Lett.} \textbf{\bibinfo{volume}{119}},
  \bibinfo{pages}{037201} (\bibinfo{year}{2017}).

\bibitem[{\citenamefont{Janssen et~al.}(2016)\citenamefont{Janssen, Andrade,
  and Vojta}}]{PhysRevLett.117.277202}
\bibinfo{author}{\bibfnamefont{L.}~\bibnamefont{Janssen}},
  \bibinfo{author}{\bibfnamefont{E.~C.} \bibnamefont{Andrade}},
  \bibnamefont{and} \bibinfo{author}{\bibfnamefont{M.}~\bibnamefont{Vojta}},
  \bibinfo{journal}{Phys. Rev. Lett.} \textbf{\bibinfo{volume}{117}},
  \bibinfo{pages}{277202} (\bibinfo{year}{2016}).

\bibitem[{\citenamefont{Winter et~al.}(2018)\citenamefont{Winter, Riedl, Kaib,
  Coldea, and Valent{\'{i}}}}]{Winter17b}
\bibinfo{author}{\bibfnamefont{S.~M.} \bibnamefont{Winter}},
  \bibinfo{author}{\bibfnamefont{K.}~\bibnamefont{Riedl}},
  \bibinfo{author}{\bibfnamefont{D.}~\bibnamefont{Kaib}},
  \bibinfo{author}{\bibfnamefont{R.}~\bibnamefont{Coldea}}, \bibnamefont{and}
  \bibinfo{author}{\bibfnamefont{R.}~\bibnamefont{Valent{\'{i}}}},
  \bibinfo{journal}{Phys. Rev. Lett.} \textbf{\bibinfo{volume}{120}},
  \bibinfo{pages}{077203} (\bibinfo{year}{2018}).

\bibitem[{\citenamefont{Wang et~al.}(2017)\citenamefont{Wang, Guo, Tafti, Hegg,
  Sen, Sidorov, Wang, Cai, Yi, Zhou et~al.}}]{Wang17b}
\bibinfo{author}{\bibfnamefont{Z.}~\bibnamefont{Wang}},
  \bibinfo{author}{\bibfnamefont{J.}~\bibnamefont{Guo}},
  \bibinfo{author}{\bibfnamefont{F.~F.} \bibnamefont{Tafti}},
  \bibinfo{author}{\bibfnamefont{A.}~\bibnamefont{Hegg}},
  \bibinfo{author}{\bibfnamefont{S.}~\bibnamefont{Sen}},
  \bibinfo{author}{\bibfnamefont{V.~A.} \bibnamefont{Sidorov}},
  \bibinfo{author}{\bibfnamefont{L.}~\bibnamefont{Wang}},
  \bibinfo{author}{\bibfnamefont{S.}~\bibnamefont{Cai}},
  \bibinfo{author}{\bibfnamefont{W.}~\bibnamefont{Yi}},
  \bibinfo{author}{\bibfnamefont{Y.}~\bibnamefont{Zhou}}, \bibnamefont{et~al.},
  \bibinfo{journal}{arXiv:1705.06139}  (\bibinfo{year}{2017}).

\bibitem[{\citenamefont{Cui et~al.}(2017)\citenamefont{Cui, Zheng, Ran, Wen,
  Liu, Liu, Guo, and Yu}}]{Cui17}
\bibinfo{author}{\bibfnamefont{Y.}~\bibnamefont{Cui}},
  \bibinfo{author}{\bibfnamefont{J.}~\bibnamefont{Zheng}},
  \bibinfo{author}{\bibfnamefont{K.}~\bibnamefont{Ran}},
  \bibinfo{author}{\bibfnamefont{J.}~\bibnamefont{Wen}},
  \bibinfo{author}{\bibfnamefont{Z.-X.} \bibnamefont{Liu}},
  \bibinfo{author}{\bibfnamefont{B.}~\bibnamefont{Liu}},
  \bibinfo{author}{\bibfnamefont{W.}~\bibnamefont{Guo}}, \bibnamefont{and}
  \bibinfo{author}{\bibfnamefont{W.}~\bibnamefont{Yu}}, \bibinfo{journal}{Phys.
  Rev. B} \textbf{\bibinfo{volume}{96}}, \bibinfo{pages}{205147}
  (\bibinfo{year}{2017}).

\bibitem[{\citenamefont{Hermann et~al.}(2018)\citenamefont{Hermann, Altmeyer,
  Ebad-Allah, Freund, Jesche, Tsirlin, Hanfland, Gegenwart, Mazin, Khomskii
  et~al.}}]{Hermann18}
\bibinfo{author}{\bibfnamefont{V.}~\bibnamefont{Hermann}},
  \bibinfo{author}{\bibfnamefont{M.}~\bibnamefont{Altmeyer}},
  \bibinfo{author}{\bibfnamefont{J.}~\bibnamefont{Ebad-Allah}},
  \bibinfo{author}{\bibfnamefont{F.}~\bibnamefont{Freund}},
  \bibinfo{author}{\bibfnamefont{A.}~\bibnamefont{Jesche}},
  \bibinfo{author}{\bibfnamefont{A.~A.} \bibnamefont{Tsirlin}},
  \bibinfo{author}{\bibfnamefont{M.}~\bibnamefont{Hanfland}},
  \bibinfo{author}{\bibfnamefont{P.}~\bibnamefont{Gegenwart}},
  \bibinfo{author}{\bibfnamefont{I.~I.} \bibnamefont{Mazin}},
  \bibinfo{author}{\bibfnamefont{D.~I.} \bibnamefont{Khomskii}},
  \bibnamefont{et~al.}, \bibinfo{journal}{Phys. Rev. B}
  \textbf{\bibinfo{volume}{97}}, \bibinfo{pages}{020104}
  (\bibinfo{year}{2018}).

\bibitem[{\citenamefont{Streltsov and Khomskii}(2017)}]{streltsov2017orbital}
\bibinfo{author}{\bibfnamefont{S.~V.} \bibnamefont{Streltsov}}
  \bibnamefont{and} \bibinfo{author}{\bibfnamefont{D.~I.}
  \bibnamefont{Khomskii}}, \bibinfo{journal}{Physics-Uspekhi}
  \textbf{\bibinfo{volume}{60}}, \bibinfo{pages}{1121} (\bibinfo{year}{2017}).

\bibitem[{\citenamefont{Veiga et~al.}(2017)\citenamefont{Veiga, Etter,
  Glazyrin, Sun, Escanhoela, Fabbris, Mardegan, Malavi, Deng, Stavropoulos
  et~al.}}]{Veiga17}
\bibinfo{author}{\bibfnamefont{L.~S.~I.} \bibnamefont{Veiga}},
  \bibinfo{author}{\bibfnamefont{M.}~\bibnamefont{Etter}},
  \bibinfo{author}{\bibfnamefont{K.}~\bibnamefont{Glazyrin}},
  \bibinfo{author}{\bibfnamefont{F.}~\bibnamefont{Sun}},
  \bibinfo{author}{\bibfnamefont{C.~A.} \bibnamefont{Escanhoela}},
  \bibinfo{author}{\bibfnamefont{G.}~\bibnamefont{Fabbris}},
  \bibinfo{author}{\bibfnamefont{J.~R.~L.} \bibnamefont{Mardegan}},
  \bibinfo{author}{\bibfnamefont{P.~S.} \bibnamefont{Malavi}},
  \bibinfo{author}{\bibfnamefont{Y.}~\bibnamefont{Deng}},
  \bibinfo{author}{\bibfnamefont{P.~P.} \bibnamefont{Stavropoulos}},
  \bibnamefont{et~al.}, \bibinfo{journal}{Phys. Rev. B}
  \textbf{\bibinfo{volume}{96}}, \bibinfo{pages}{140402}
  (\bibinfo{year}{2017}).

\bibitem[{\citenamefont{Hentrich et~al.}(2017)\citenamefont{Hentrich, Wolter,
  Zotos, Brenig, Nowak, Isaeva, Doert, Banerjee, Lampen-Kelley, Mandrus
  et~al.}}]{Hentrich17}
\bibinfo{author}{\bibfnamefont{R.}~\bibnamefont{Hentrich}},
  \bibinfo{author}{\bibfnamefont{A.~U.~B.} \bibnamefont{Wolter}},
  \bibinfo{author}{\bibfnamefont{X.}~\bibnamefont{Zotos}},
  \bibinfo{author}{\bibfnamefont{W.}~\bibnamefont{Brenig}},
  \bibinfo{author}{\bibfnamefont{D.}~\bibnamefont{Nowak}},
  \bibinfo{author}{\bibfnamefont{A.}~\bibnamefont{Isaeva}},
  \bibinfo{author}{\bibfnamefont{T.}~\bibnamefont{Doert}},
  \bibinfo{author}{\bibfnamefont{A.}~\bibnamefont{Banerjee}},
  \bibinfo{author}{\bibfnamefont{P.}~\bibnamefont{Lampen-Kelley}},
  \bibinfo{author}{\bibfnamefont{D.~G.} \bibnamefont{Mandrus}},
  \bibnamefont{et~al.}, \bibinfo{journal}{arXiv:1703.08623}
  (\bibinfo{year}{2017}).

\bibitem[{\citenamefont{Beyer and Dressel}(2015)}]{Beyer15}
\bibinfo{author}{\bibfnamefont{R.}~\bibnamefont{Beyer}} \bibnamefont{and}
  \bibinfo{author}{\bibfnamefont{M.}~\bibnamefont{Dressel}},
  \bibinfo{journal}{Rev. Sci. Instrum.} \textbf{\bibinfo{volume}{86}},
  \bibinfo{pages}{053904} (\bibinfo{year}{2015}).

\bibitem[{\citenamefont{Hasegawa et~al.}(2017)\citenamefont{Hasegawa, Aoyama,
  Sasaki, Ikemoto, Moriwaki, Shirakura, Saito, Imai, and Ohgushi}}]{Hasegawa17}
\bibinfo{author}{\bibfnamefont{Y.}~\bibnamefont{Hasegawa}},
  \bibinfo{author}{\bibfnamefont{T.}~\bibnamefont{Aoyama}},
  \bibinfo{author}{\bibfnamefont{K.}~\bibnamefont{Sasaki}},
  \bibinfo{author}{\bibfnamefont{Y.}~\bibnamefont{Ikemoto}},
  \bibinfo{author}{\bibfnamefont{T.}~\bibnamefont{Moriwaki}},
  \bibinfo{author}{\bibfnamefont{T.}~\bibnamefont{Shirakura}},
  \bibinfo{author}{\bibfnamefont{R.}~\bibnamefont{Saito}},
  \bibinfo{author}{\bibfnamefont{Y.}~\bibnamefont{Imai}}, \bibnamefont{and}
  \bibinfo{author}{\bibfnamefont{K.}~\bibnamefont{Ohgushi}},
  \bibinfo{journal}{J. Phys. Soc. Japan} \textbf{\bibinfo{volume}{86}},
  \bibinfo{pages}{123709} (\bibinfo{year}{2017}).

\bibitem[{\citenamefont{Glamazda et~al.}(2017)\citenamefont{Glamazda, Lemmens,
  Do, Kwon, and Choi}}]{Glamazda17}
\bibinfo{author}{\bibfnamefont{A.}~\bibnamefont{Glamazda}},
  \bibinfo{author}{\bibfnamefont{P.}~\bibnamefont{Lemmens}},
  \bibinfo{author}{\bibfnamefont{S.-H.} \bibnamefont{Do}},
  \bibinfo{author}{\bibfnamefont{Y.~S.} \bibnamefont{Kwon}}, \bibnamefont{and}
  \bibinfo{author}{\bibfnamefont{K.-Y.} \bibnamefont{Choi}},
  \bibinfo{journal}{Phys. Rev. B} \textbf{\bibinfo{volume}{95}},
  \bibinfo{pages}{174429} (\bibinfo{year}{2017}).

\bibitem[{\citenamefont{Giannozzi et~al.}(2009)\citenamefont{Giannozzi, Baroni,
  Bonini, Calandra, Car, Cavazzoni, Ceresoli, Chiarotti, Cococcioni, Dabo
  et~al.}}]{Giannozzi09}
\bibinfo{author}{\bibfnamefont{P.}~\bibnamefont{Giannozzi}},
  \bibinfo{author}{\bibfnamefont{S.}~\bibnamefont{Baroni}},
  \bibinfo{author}{\bibfnamefont{N.}~\bibnamefont{Bonini}},
  \bibinfo{author}{\bibfnamefont{M.}~\bibnamefont{Calandra}},
  \bibinfo{author}{\bibfnamefont{R.}~\bibnamefont{Car}},
  \bibinfo{author}{\bibfnamefont{C.}~\bibnamefont{Cavazzoni}},
  \bibinfo{author}{\bibfnamefont{D.}~\bibnamefont{Ceresoli}},
  \bibinfo{author}{\bibfnamefont{G.~L.} \bibnamefont{Chiarotti}},
  \bibinfo{author}{\bibfnamefont{M.}~\bibnamefont{Cococcioni}},
  \bibinfo{author}{\bibfnamefont{I.}~\bibnamefont{Dabo}}, \bibnamefont{et~al.},
  \bibinfo{journal}{J. Phys. Condens. Matter} \textbf{\bibinfo{volume}{21}},
  \bibinfo{pages}{395502} (\bibinfo{year}{2009}).

\bibitem[{\citenamefont{Baroni et~al.}(2001)\citenamefont{Baroni, de~Gironcoli,
  {Dal Corso}, and Giannozzi}}]{Baroni01}
\bibinfo{author}{\bibfnamefont{S.}~\bibnamefont{Baroni}},
  \bibinfo{author}{\bibfnamefont{S.}~\bibnamefont{de~Gironcoli}},
  \bibinfo{author}{\bibfnamefont{A.}~\bibnamefont{{Dal Corso}}},
  \bibnamefont{and}
  \bibinfo{author}{\bibfnamefont{P.}~\bibnamefont{Giannozzi}},
  \bibinfo{journal}{Rev. Mod. Phys.} \textbf{\bibinfo{volume}{73}},
  \bibinfo{pages}{515} (\bibinfo{year}{2001}).

\bibitem[{SM()}]{SM}
\bibinfo{note}{See Supplemental Material at http://link.aps.org./ supplemental/
  for more details.}

\bibitem[{\citenamefont{Fletcher et~al.}(1967)\citenamefont{Fletcher, Gardner,
  Fox, and Topping}}]{Fletcher67}
\bibinfo{author}{\bibfnamefont{J.~M.} \bibnamefont{Fletcher}},
  \bibinfo{author}{\bibfnamefont{W.~E.} \bibnamefont{Gardner}},
  \bibinfo{author}{\bibfnamefont{A.~C.} \bibnamefont{Fox}}, \bibnamefont{and}
  \bibinfo{author}{\bibfnamefont{G.}~\bibnamefont{Topping}},
  \bibinfo{journal}{J. Chem. Soc. A} p. \bibinfo{pages}{1038}
  (\bibinfo{year}{1967}).

\bibitem[{\citenamefont{Binotto et~al.}(1971)\citenamefont{Binotto, Pollini,
  and Spinolo}}]{Binotto71}
\bibinfo{author}{\bibfnamefont{L.}~\bibnamefont{Binotto}},
  \bibinfo{author}{\bibfnamefont{I.}~\bibnamefont{Pollini}}, \bibnamefont{and}
  \bibinfo{author}{\bibfnamefont{G.}~\bibnamefont{Spinolo}},
  \bibinfo{journal}{Phys. Stat. Sol. (b)} \textbf{\bibinfo{volume}{44}},
  \bibinfo{pages}{245} (\bibinfo{year}{1971}).

\bibitem[{\citenamefont{Guizzetti et~al.}(1979)\citenamefont{Guizzetti,
  Reguzzoni, and Pollini}}]{Guizzetti79}
\bibinfo{author}{\bibfnamefont{G.}~\bibnamefont{Guizzetti}},
  \bibinfo{author}{\bibfnamefont{E.}~\bibnamefont{Reguzzoni}},
  \bibnamefont{and} \bibinfo{author}{\bibfnamefont{I.}~\bibnamefont{Pollini}},
  \bibinfo{journal}{Phys. Lett. A} \textbf{\bibinfo{volume}{70}},
  \bibinfo{pages}{34} (\bibinfo{year}{1979}).

\bibitem[{\citenamefont{Sandilands
  et~al.}(2016{\natexlab{a}})\citenamefont{Sandilands, Sohn, Park, Kim, Kim,
  Sears, Kim, and Noh}}]{Sandilands16b}
\bibinfo{author}{\bibfnamefont{L.~J.} \bibnamefont{Sandilands}},
  \bibinfo{author}{\bibfnamefont{C.~H.} \bibnamefont{Sohn}},
  \bibinfo{author}{\bibfnamefont{H.~J.} \bibnamefont{Park}},
  \bibinfo{author}{\bibfnamefont{S.~Y.} \bibnamefont{Kim}},
  \bibinfo{author}{\bibfnamefont{K.~W.} \bibnamefont{Kim}},
  \bibinfo{author}{\bibfnamefont{J.~A.} \bibnamefont{Sears}},
  \bibinfo{author}{\bibfnamefont{Y.-J.} \bibnamefont{Kim}}, \bibnamefont{and}
  \bibinfo{author}{\bibfnamefont{T.~W.} \bibnamefont{Noh}},
  \bibinfo{journal}{Phys. Rev. B} \textbf{\bibinfo{volume}{94}},
  \bibinfo{pages}{195156} (\bibinfo{year}{2016}{\natexlab{a}}).

\bibitem[{\citenamefont{Sandilands
  et~al.}(2016{\natexlab{b}})\citenamefont{Sandilands, Tian, Reijnders, Kim,
  Plumb, Kim, Kee, and Burch}}]{Sandilands16a}
\bibinfo{author}{\bibfnamefont{L.~J.} \bibnamefont{Sandilands}},
  \bibinfo{author}{\bibfnamefont{Y.}~\bibnamefont{Tian}},
  \bibinfo{author}{\bibfnamefont{A.~A.} \bibnamefont{Reijnders}},
  \bibinfo{author}{\bibfnamefont{H.-S.} \bibnamefont{Kim}},
  \bibinfo{author}{\bibfnamefont{K.~W.} \bibnamefont{Plumb}},
  \bibinfo{author}{\bibfnamefont{Y.-J.} \bibnamefont{Kim}},
  \bibinfo{author}{\bibfnamefont{H.-Y.} \bibnamefont{Kee}}, \bibnamefont{and}
  \bibinfo{author}{\bibfnamefont{K.~S.} \bibnamefont{Burch}},
  \bibinfo{journal}{Phys. Rev. B} \textbf{\bibinfo{volume}{93}},
  \bibinfo{pages}{075144} (\bibinfo{year}{2016}{\natexlab{b}}).

\bibitem[{\citenamefont{Koitzsch et~al.}(2017)\citenamefont{Koitzsch,
  M{\"u}ller, Knupfer, B{\"u}chner, Nowak, Isaeva, Doert, Gr{\"u}ninger,
  Nishimoto, and van~den Brink}}]{Koitzsch17b}
\bibinfo{author}{\bibfnamefont{A.}~\bibnamefont{Koitzsch}},
  \bibinfo{author}{\bibfnamefont{E.}~\bibnamefont{M{\"u}ller}},
  \bibinfo{author}{\bibfnamefont{M.}~\bibnamefont{Knupfer}},
  \bibinfo{author}{\bibfnamefont{B.}~\bibnamefont{B{\"u}chner}},
  \bibinfo{author}{\bibfnamefont{D.}~\bibnamefont{Nowak}},
  \bibinfo{author}{\bibfnamefont{A.}~\bibnamefont{Isaeva}},
  \bibinfo{author}{\bibfnamefont{T.}~\bibnamefont{Doert}},
  \bibinfo{author}{\bibfnamefont{M.}~\bibnamefont{Gr{\"u}ninger}},
  \bibinfo{author}{\bibfnamefont{S.}~\bibnamefont{Nishimoto}},
  \bibnamefont{and} \bibinfo{author}{\bibfnamefont{J.}~\bibnamefont{van~den
  Brink}}, \bibinfo{journal}{arXiv:1709.02712}  (\bibinfo{year}{2017}).

\bibitem[{\citenamefont{Khaliullin et~al.}(2004)\citenamefont{Khaliullin,
  Horsch, and Ole{\'{s}}}}]{Khaliullin04}
\bibinfo{author}{\bibfnamefont{G.}~\bibnamefont{Khaliullin}},
  \bibinfo{author}{\bibfnamefont{P.}~\bibnamefont{Horsch}}, \bibnamefont{and}
  \bibinfo{author}{\bibfnamefont{A.~M.} \bibnamefont{Ole{\'{s}}}},
  \bibinfo{journal}{Phys. Rev. B} \textbf{\bibinfo{volume}{70}},
  \bibinfo{pages}{195103} (\bibinfo{year}{2004}).

\bibitem[{\citenamefont{Rau et~al.}(2014)\citenamefont{Rau, Lee, and
  Kee}}]{Rau14}
\bibinfo{author}{\bibfnamefont{J.~G.} \bibnamefont{Rau}},
  \bibinfo{author}{\bibfnamefont{E.~K.-H.} \bibnamefont{Lee}},
  \bibnamefont{and} \bibinfo{author}{\bibfnamefont{H.-Y.} \bibnamefont{Kee}},
  \bibinfo{journal}{Phys. Rev. Lett.} \textbf{\bibinfo{volume}{112}},
  \bibinfo{pages}{077204} (\bibinfo{year}{2014}).

\bibitem[{\citenamefont{Winter et~al.}(2016)\citenamefont{Winter, Li, Jeschke,
  and Valent{\'{i}}}}]{Winter16}
\bibinfo{author}{\bibfnamefont{S.~M.} \bibnamefont{Winter}},
  \bibinfo{author}{\bibfnamefont{Y.}~\bibnamefont{Li}},
  \bibinfo{author}{\bibfnamefont{H.~O.} \bibnamefont{Jeschke}},
  \bibnamefont{and}
  \bibinfo{author}{\bibfnamefont{R.}~\bibnamefont{Valent{\'{i}}}},
  \bibinfo{journal}{Phys. Rev. B} \textbf{\bibinfo{volume}{93}},
  \bibinfo{pages}{214431} (\bibinfo{year}{2016}).

\end{thebibliography}

\end{document}